\documentclass[12pt]{iopart}
\usepackage{graphicx}% Include figure files

\def\bra#1{\mathinner{\langle{#1}|}}
\def\ket#1{\mathinner{|{#1}\rangle}}
\begin{document}

\title[Charge relaxation]{Mesoscopic capacitance oscillations}

\author{Markus B\"uttiker and Simon E. Nigg}

\address{D\'epartement de Physique Th\'eorique, Universit\'e de
  Gen\`eve, CH-1211 Gen\`eve 4, Switzerland}
\ead{Markus.Buttiker@physics.unige.ch}
\begin{abstract}
We examine oscillations as a function of Fermi energy in the capacitance of a mesoscopic cavity connected via a single quantum channel to a metallic contact and capacitively coupled to a gate. The oscillations depend on the distribution of single levels in the cavity, the interaction strength and the transmission probability through the quantum channel. We use a Hartree-Fock approach to exclude self-interaction. The sample specific capacitance oscillations are in marked contrast to the charge relaxation resistance, which together with the capacitance defines the RC-time, and which for spin polarized electrons is quantized at half a resistance quantum. Both the capacitance oscillations and the quantized charge relaxation resistance are seen in a strikingly clear manner in a recent experiment. 
\end{abstract}

\maketitle

\section{Capacitance and charge relaxation}

\begin{figure}
\begin{center}
\includegraphics[width=0.45\textwidth]{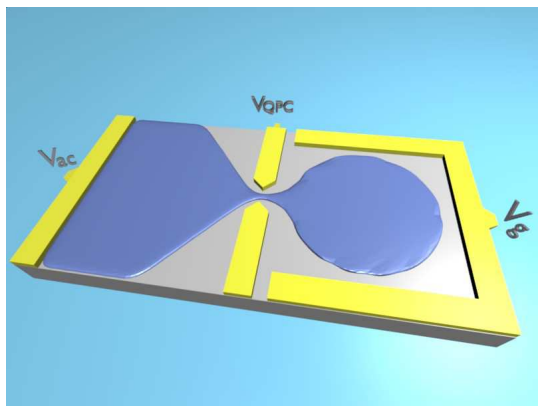}
\caption{\label{fig:dot} (color online). Mesoscopic capacitor: A
 cavity is connected via one lead to an electron reservoir at voltage
 $V_{ac}$ and capacitively coupled to a gate with voltage
 $V_g$. Only one contact formed by a quantum point contact permits carrier exchange.   The voltage $V_{qpc}$ controlls the transmission through the quantum point contact.}
 \end{center}
\end{figure}

We are interested in the dynamics of a quantum coherent capacitor. We ask: if an excess charge has been created due to a sudden change in applied voltage or due to a spontaneous fluctuation, how long does it take for this excess charge to relax? For a classical capacitor the answer is 
simply the $RC-$time. Even for a quantum coherent capacitor, the answer can be expressed
in terms of a capacitance and a charge relaxation resistance. However, the properties of these two transport coefficients can now differ dramatically from their classical counterparts.
We use the term {\it electrochemical capacitance} to emphasize that 
the capacitance now depends on the physical properties of the system, and use the term {\it charge relaxation resistance} to emphasize that the resistance might be very different from what is expected classically. 

The system of interest is shown in figure 1. A small cavity is connected via a single lead 
to a metallic contact and is capacitively coupled to a gate at voltage $V_g$. The cavity acts as one plate of a capacitor and the gate as the second plate.  The opening between the cavity 
and the metallic contact is formed by a quantum point contact
(QPC). Of interest is the entire transition from very weak coupling to
very strong coupling when electrons can transmit the QPC with
transmission probability 1. A structure of this type was analyzed more
than 13 years ago, by B\"uttiker, Pr\^etre and Thomas
\cite{Buettiker:93a} who found a charge relaxation resistance which in
the case of a single channel of spin polarized electrons is given by
half a resistance quantum independent of the transmission probability
of the channel. Interestingly, a recent experiment by Gabelli et
al. \cite{Gabelli} on a structure similar to that shown in figure 1 measured both the in-phase and the out-of-phase admittance and found excellent agreement with the 
predictions of Ref.  \cite{Buettiker:93a}. In particular their experiments confirm the quantization of the charge relaxation resistance. 

Much of mesoscopic physics has centered around a few elementary systems which best demonstrate 
a quantum effect of interest. Examples are isolated rings which were shown to support a persistent current \cite{bil,exp1,exp2,exp3}, rings with leads exhibit an Aharonov-Bohm effect in conductance \cite{gia,bia,webb,hansen}, QPC's \cite{qpc1,qpc2} and quantum Hall bars \cite{butt88} teach us about conductance quantization and quantum dots \cite{kouwen} reveal in a striking manner Coulomb blockade and Kondo physics. Clearly the mesoscopic capacitor is also such an elementary system: its major advantage is a complete suppression of dc-transport. Unlike in systems with dc-transport where very high frequencies are required to make dynamic effects visible on top of the zero-frequency dc-response, here the dynamic effects provide the leading order of the system response.  

\section{The admittance of a capacitor}

Classically one might want to consider the sample in figure 1 as a series connection of a geometrical capacitance $C$ and a resistance $R$. The capacitance $C$ is the geometric capacitance between the cavity and the gate and the resistance $R$ is that of the QPC. At low frequencies the conductance $G(\omega) \equiv dI(\omega)/dV(\omega)$ can be expanded in a power series to give 
\begin{equation}\label{eq:0}
G(\omega) = -i\omega C +\omega^2 C^2 R + O(\omega^3).
\end{equation}
The first term describes an out-of-phase response determined by the
geometrical capacitance $C$, whereas the second term is the in-phase, dissipative response determined by the resistance $R$.
The quantum coherent capacitor of interest here has an ac-conductance which can similarly be expanded in powers of $\omega$ but instead of the classical quantities
$C$ and $R$ is determined by an electrochemical capacitance $C_{\mu}$ and a charge relaxation resistance $R_q$,
\begin{equation}\label{eq:1}
G(\omega) = -i\omega C_{\mu}+\omega^2 C_{\mu}^2 R_q+O(\omega^3).
\end{equation}
The electrochemical capacitance $C_{\mu}$ differs from $C$ whenever the 
small system is not efficient in screening electric fields. The charge relaxation resistance $R_q$  reflects the fact that carriers injected into the cavity spend a time on this "plate" which is too short to equilibrate them \cite{Buettiker:93a,Buettiker:96a} and for this reason $R_q$ might differ from $R$.  The capacitance $C_{\mu}$
is related to the imaginary part of the AC conductance \cite{Buettiker:93a,Buettiker:96a,pomorski} and is often measured dynamically \cite{Gabelli,Sillanpaa:05a,Duty} but in theory it can also
be obtained by differentiation of a thermodynamic
(grand-canonical) potential
\cite{Flensberg:93a,Matveev:95a,lehur1,Buettiker:96b}. 
Here we pursue a dynamic approach. 

B\"uttiker, Thomas and Pr\^etre \cite{Buettiker:93a} developed a theory of charge relaxation of quantum coherent capacitors based on scattering theory and a self-consistent (Hartree like) treatment of interactions. Carriers from the reservoir enter the cavity and are eventually reflected. 
If the contact supports $N$ transverse channels, the reflection amplitudes can be described 
by an $N\times N$ scattering matrix $S$. Since all carriers are reflected the eigen values of this scattering matrix are of the form $\exp{i\phi_n}$ where $\phi_n$ is the increment in phase
which a particle in the n-th eigen channel acquires upon reflection.
Both $C_{\mu}$
and $R_q$ depend on the derivative of such phases with respect to energy $d\phi_{n} (E)/dE$.
The derivative is taken at the Fermi energy. When multiplied by $\hbar$ the phase derivatives
are the Wigner-Smith delay times. Alternatively we can say that 

\begin{equation}\label{eq:2} 
\nu_n (E)= (1/2\pi) d\phi_{n} (E)/dE
\end{equation}
is the contribution to the density of states in the cavity of the n-th eigen channel. 
The total density of states in the cavity at energy $E$ is $\nu(E) = (1/2\pi) Tr(S^{\dagger}dS/dE)= \sum_{n} \nu_n (E)$. Assuming that the potential inside the cavity can be described by a single variable and that Coulomb coupling to the gate is described by a geometrical capacitance $C$, 
Refs. \cite{Buettiker:93a,Buettiker:96a} find for the electrochemical capacitance 

\begin{equation}\label{eq:3} 
C_{\mu} = \frac{C e^{2}\nu(E)}{C + e^{2}\nu(E)} \,.
\end{equation}
If the Coulomb interaction is weak $C \gg e^{2}\nu(E)$ the electrochemical capacitance is determined by the density of states $C_{\mu} \approx e^{2}\nu(E)$, whereas if the energy to charge the cavity is large the capacitance is essentially determined by the geometrical capacitance $C_{\mu} \approx C$. 
Due to quantum interference the density of states $\nu(E)$ exhibits mesoscopic fluctuations from 
sample to sample: no two capacitances are exactly equal. For a single channel connected to a chaotic cavity these fluctuations are discussed by Gopar, Mello and B\"uttiker \cite{gopar}. For a cavity coupled to many channel QPC's mesoscopic capacitance fluctuations are treated by Brouwer and B\"uttiker \cite{brouwer1} and Brouwer et al.\cite{brouwer2}. 

For the charge relaxation resistance Ref. \cite{Buettiker:93a} finds, 

\begin{equation}\label{eq:4} 
R_q = \frac{h}{2e^2}\frac{\sum_n \nu^{2}_n (E)}{[\sum_n \nu_n (E)]^2 }. 
\end{equation}
Perhaps the most striking prediction of this approach is that if the connection between cavity and reservoir permits transmission of at most one spin polarized quantum channel (n =1 in equation (\ref{eq:4})), 
the charge relaxation resistance  at
zero temperature is equal to half a resistance quantum~\cite{Buettiker:93a,Buettiker:96a}

\begin{equation}\label{eq:5}
R_q = \frac{h}{2e^2}.
\end{equation}
This should be contrasted with the resistance $R \propto \frac{h}{e^2} \frac{1}{T}$ with $T$ the transmission probability, which we would expect if inelastic scattering is sufficiently strong to relax the carriers in the cavity toward 
an equilibrium distribution. The prediction is for a resistance that is entirely independent of the scattering properties of the channel! Note that the predicted charge relaxation resistance is 
not ${h}/{e^2}$, the resistance of a perfect two-probe single channel conductor but only half a dc-resistance quantum of a single channel.

The experiment by Gabelli et al.~\cite{Gabelli} measures both the in and out of phase parts of the AC conductance of a mesoscopic RC circuit. The sample similar to figure 1 with a sub-micron sized 
cavity is cooled to sub-Kelvin temperatures and subjected to an ac voltage in the GHz range. A high magnetic field is applied such that only a single spin polarized edge channel is partially transmitted and reflected at the QPC. Gabelli et al.~\cite{Gabelli} develop a model 
which permits to determine $C_{\mu}$ and $R_q$ using Eqs. (\ref{eq:3},\ref{eq:4})
and compare with experiment. In their model the scattering matrix follows from Fabry-P{\'e}rot 
like multiple reflections in the cavity. 

The results of this experiment
are in good agreement with the theoretical predictions of~\cite{Buettiker:93a}, in particular they confirm 
the universality of the single channel charge relaxation resistance over a range of voltages 
over which the transmission probability of the QPC varies substantially.  
Two effects might give rise to deviations from 
the quantized charge relaxation resistance, equation (\ref{eq:5}): First, as the transmission probability $T$ through the contact becomes sufficiently small, the time a carrier spends inside the cavity will eventually become long compared to an inelastic time. As a consequence we expect realistically that for very
small contact transparencies there is a cross-over from $R_q$ given by equation (\ref{eq:5}) to the "classical" result $R \propto \frac{h}{e^2} \frac{1}{T} \gg \frac{h}{e^2}$. Second, for small transparencies, charge quantization (Coulomb blockade) will become important and the random phase approach used by 
B\"uttiker et al.\cite{Buettiker:93a} might fail. The second question has already been addressed in recent work by Nigg, Lopez and B\"uttiker \cite{nigg}. These authors show that the charge relaxation resistance $R_q$ remains quantized even in the strong Coulomb blockade limit. 
To demonstrate this, an approach is developed which is based on a Hartree-Fock treatment which explicitly excludes self-interaction. For many channel contacts, the resulting expressions for $R_q$, like the expressions derived from the Hartree approach, can be formulated in terms of density of states or the time-delay of particles. 

It is the purpose of this work to present the results for the electrochemical capacitance $C_{\mu}$
in the presence of interactions treated on the Hartree-Fock level. Below we outline how the density of states are obtained within a Hartree-Fock approach. We then present results for the mesoscopic capacitance and the total charge in the dot as the transmission through the contact increases from zero to one.

\section{Density of states and charge steps}

For a moment consider the cavity disconnected from the lead.  In Hartree-Fock the isolated dot is described by the effective single particle Hamiltonian $H$. In the eigen basis $\{\ket{{\lambda}}\}$ of the Hamiltonian, we have
$H=\sum_{\lambda}E_{\lambda}\ket{{\lambda}}\bra{{\lambda}}$ where $E_{\lambda}$ are the Hartree-Fock level energies. Coupling the cavity via a single quantum channel to the states in the leads with 
tunnel matrix elements $W_{\lambda}$ gives rise to a scattering matrix 
%and $W = \sum_{\lambda}W_{\lambda}\ket{\lambda}\bra{\lambda}$ and (\ref{eq:12})
%becomes
~\cite{wigner}
\begin{equation}\label{eq:6}
S(E) = \frac{1 + iK(E)}{1 - iK(E)},
\end{equation}
with
\begin{equation}\label{eq:7}
K(E) =
\sum_{\lambda}\frac{\Gamma_{\lambda}}{E_{\lambda}-E}
\end{equation}
where
\begin{equation}\label{eq:8}
\Gamma_{\lambda}=\pi W_{\lambda}^*W_{\lambda}.
\end{equation}
The function $K(E)$ determines the phase $\phi(E)$ which a carrier acquires upon reflection through $\phi = 2 \arctan(K(E))$. 
The density of states of the cavity $\nu(E) = (1/2\pi i) S^*(E)dS(E)/dE = (1/\pi)d\arctan(K(E))/dE$ 
is then given by $\nu(E) = \sum_{\lambda}\nu_{\lambda}(E)$
with
\begin{equation}\label{eq:9}
\nu_{\lambda}(E) = \frac{1}{\pi}\frac{\Gamma_{\lambda}}{(E_{\lambda}-E)^2+\left[(E_{\lambda}-E)\sum_{\mu}\frac{\Gamma_{\mu}}{E_{\mu}-E}\right]^2}.
\end{equation}
Note that this expression for the density of states differs from the often used Breit-Wigner expression. According to equation (\ref{eq:9}) the density of states at every energy depends on all the eigen energies (poles), except exactly at the resonance energy $E = E_{\lambda}$ where the density of states is $\nu(E_{\lambda})= {1}/{\pi} \Gamma_{\lambda}$. We next specify how we treat interaction. 
In Hartree-Fock the self-energy is \cite{Hackenbroich:01a}
\begin{equation}\label{eq:10}
(\Sigma_{HF})_{mi}= E_c\left[\delta_{mi}\sum_{l}\langle n_l \rangle - \langle d_i^{\dagger}d_m \rangle \right].
\end{equation}
Here $E_c = e^2 /C$ is a charging energy. 
The first term in equation (\ref{eq:10}) is the Hartree energy determined by the occupied states. The occupation number operator of the state $l$ in the dot is  $n_l = d_l^{\dagger}d_l $ where $d_l$ annihilates a carrier in the cavity. The second term in equation (\ref{eq:10}) is 
of central importance for the description of charge quantization effects: the second 
term excludes self-interaction. Note that the self-energy is a matrix
with dimension $M\times M$
equal to the number of states in the dot. In the following we neglect the off-diagonal elements of this matrix. These elements  account for weak inter-level
exchange interactions and for the parameters of interest here affect the results only quantitatively \cite{nigg}. With this approximation the self-consistent Hartree-Fock level energies are given by
\begin{equation}\label{eq:11}
E_{\lambda}= \epsilon_{\lambda}+E_c \sum_{\underline{\mu}\not= \lambda}\langle n_{\mu}\rangle ,
\end{equation}
where $\langle n_{\mu}\rangle =\int dE f(E)\nu_{\mu}(E)$ and $\epsilon_{\mu}$
is the bare (non-interacting) energy of level $\mu$. Eqs. (\ref{eq:9},\ref{eq:10},\ref{eq:11}) are a set of self-consistent equations which we solve iteratively. 

\begin{figure}[]
\begin{center}
\includegraphics[width=0.45\textwidth]{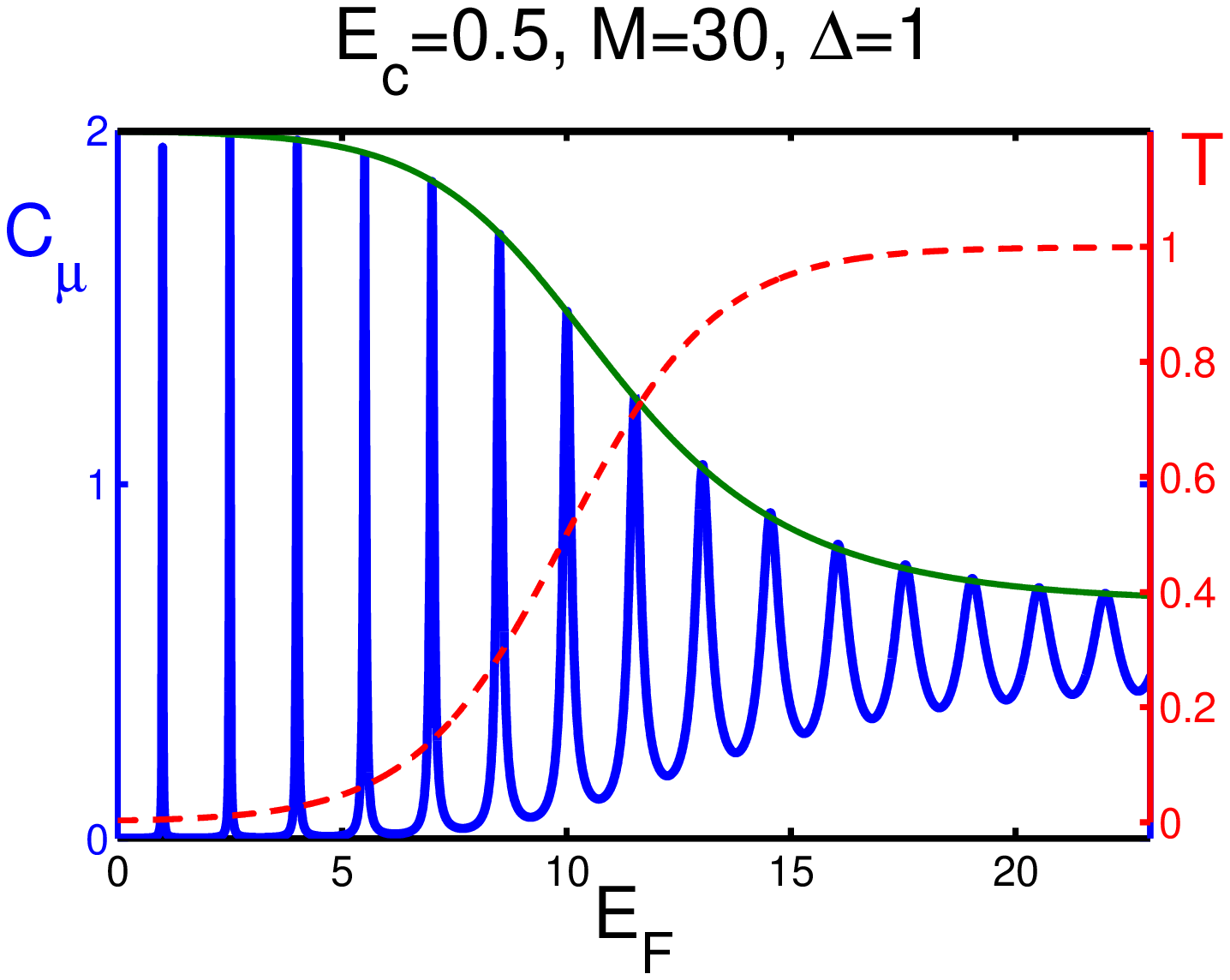}\hfill
\includegraphics[width=0.45\textwidth]{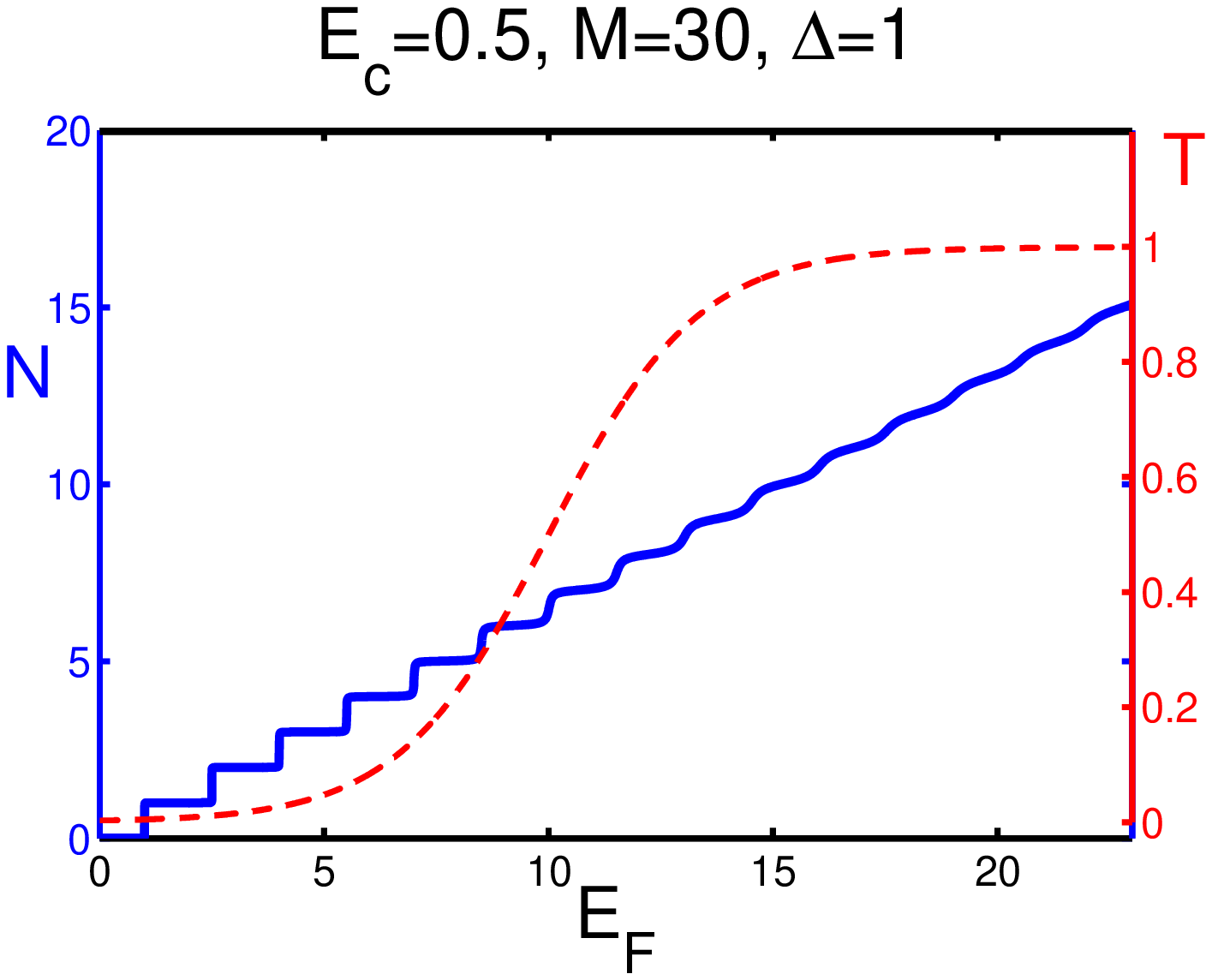}\\
\includegraphics[width=0.45\textwidth]{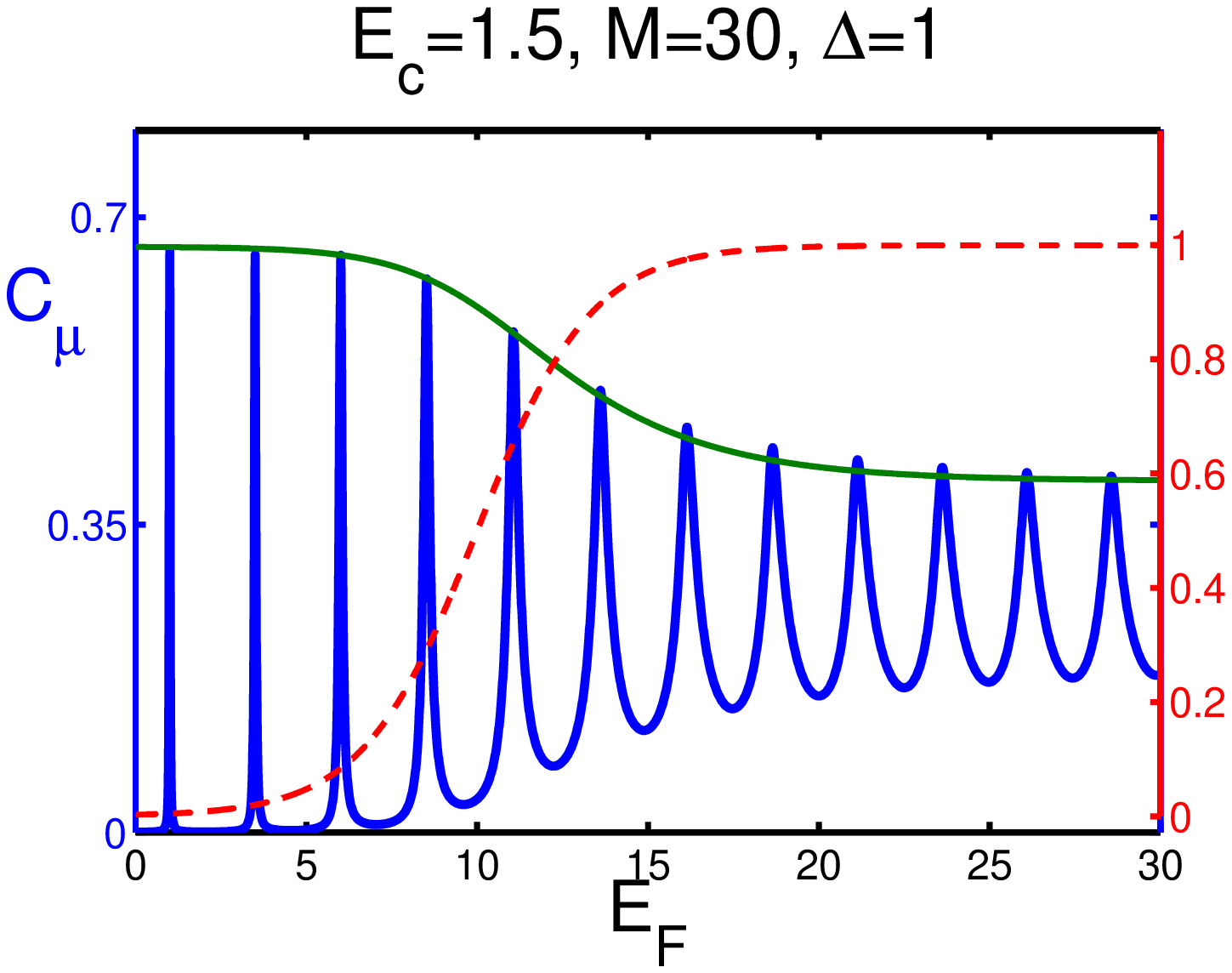}\hfill
\includegraphics[width=0.45\textwidth]{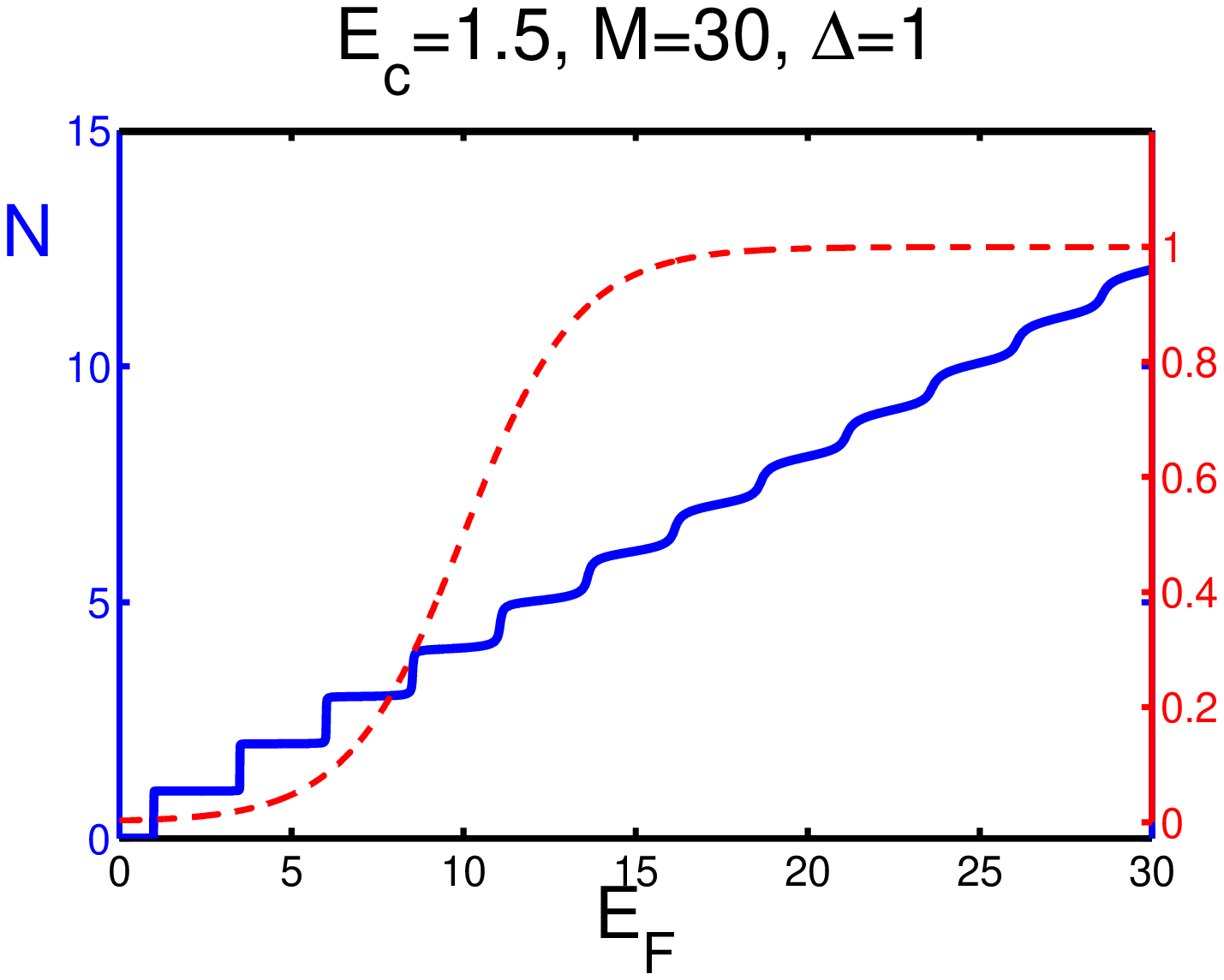}
\caption{(color online) Left: Electrochemical
capacitance oscillations (solid blue line) as a function of the Fermi
energy. The (green) curve following the capacitance peaks is the capacitance determined by the escape rate (see text). Right: Charge steps  as a function of Fermi energy. The transmission probability through the QPC is shown in all figures as a (red) dashed curve.}\label{fig:cmu}
\end{center}
\end{figure}

\section{Capacitance oscillations}

To proceed, we now assume for simplicity that the non-interacting
energy levels are equidistant in energy with a spacing $\Delta_\lambda
= \Delta$. We also assume that all levels couple with the same
strength $\Gamma_{\lambda}= \gamma$ to the lead. Furthermore we use a
result from Ref. \cite{rmt} which relates the coupling constant and the bare level spacing to the transmission probability $T$ through the contact, 
\begin{equation}\label{eq:12}
\gamma =\frac{\Delta}{\pi
  T}\left(2-T-2\sqrt{1-T}\right).
\end{equation}
This is a local relation and thus also valid in the Coulomb blockade regime. 
To complete our model we connect the transmission probability through the QPC to the Fermi energy, $T=1/[1+\exp\left\{a(E_F-E_0)\right\}]$.  This expression is appropriate for a QPC with an (inverted) parabolic potential \cite{Buttiker90}. Here $E_0$ determines the energy at which transmission is $1/2$ and $a$ is a constant which depends on the curvatures of the saddle point potential.

In figure~\ref{fig:cmu} we show the zero temperature electrochemical
capacitance $C_{\mu}$ and the
total dot charge $N$ for different interaction strengths $E_C/\Delta= 0.5$
and $1.5$ as a function of the
Fermi energy.  In all the graphs the probability for transmission through the QPC is shown as a dashed red curve. The total number of levels in the dot is $M=30$.
As expected, for very small transmission, the capacitance has sharp Coulomb peaks at energies
at which transfer of a carrier into the dot is permitted. The charge increases step like with flat plateau regions in which charge is quantized. As the contact is opened the peak height decreases and the width of the capacitive Coulomb peaks increases. Even when the quantum channel is completely open, the density of states, equation (\ref{eq:9}), exhibits weak oscillations. This is similar to a chaotic cavity which even for a completely transparent quantum channel exhibits density of states fluctuations and thus capacitance fluctuations \cite{gopar,brouwer1,brouwer2}. The existence of oscillations for $T = 1$ differs from the model used by Gabelli et al. \cite{Gabelli} 
where the cavity was treated as a Fabry-P{\'e}rot resonator. In the model of Ref. \cite{Gabelli} the  capacitance oscillations are proportional to the amplitude of reflection of the QPC. The model of Gabelli et al. \cite{Gabelli} is appropriate for electron motion along edge states. Clearly, if the QPC becomes completely transparent to an edge state there is no interference, thus no density oscillation and thus no capacitance oscillation. In contrast the approach used here is suitable if only a weak magnetic field is applied (corresponding to a flux through the cavity of the order of a flux quantum). 

The (green) curves connecting the peaks in figure 2 show the capacitance calculated with the density of states 
$\nu(E_F ) = 1/\pi \gamma (E_F) $. Deep in the Coulomb blockade regime this function coincides with the local maxima of the electrochemical capacitance, however, with increasing transparency, as neighboring peaks start to overlap, the maxima of the capacitance cease to be determined by a single Hartree-Fock level.  

The graphs on the right hand side of figure 2 show that with increasing transparency trough the QPC charge oscillations become ever weaker and for large transmission we have a nearly continuous increase of the charge in the cavity. While semi-classical theory of the Coulomb blockade leads to charge steps which are infinitely sharp in the zero-temperature limit, here 
the sharpness of the charge steps depends on the transmission probability. Other systems which 
exhibit charge steps which are not infinitely sharp even in the zero temperature limit 
are Cooper pair boxes \cite{mb87,bouchiat}(where the slope has been measured \cite{bouchiat,Sillanpaa:05a,Duty}) but also single electron systems like rings with in line or side quantum dots \cite{Buettiker:96b}. 

\section{Conclusions}

In this work we have examined Coulomb blockade oscillations for mesoscopic quantum coherent capacitors. We present an effective single particle approach using scattering theory and 
treating interactions on the level of a Hartree-Fock approach which excludes self-interactions.
This approach preserves quantum coherence and at the same time takes effects of charge quantization into account.
For a nearly isolated cavity the capacitance exhibits sharp and well seprated Coulomb peaks. As expected, with increasing coupling between cavity and reservoir charge quantization effects diminish. However, interestingly, we find that even for a completely transparent quantum channel capacitance oscillations persist. These oscillations reflect quantum coherent scattering in the cavity. 

The results presented here can be experimentally tested. Experiments \cite{Gabelli} in high magnetic fields have already reported clear signatures of a quantum coherent capacitor. Our results should be useful in finding such signatures in the low magnetic field range.

\section*{Acknowledgment}
We thank Julien Gabelli and Christian Glattli for numerous discussions of their experiment and Piet 
Brouwer for comments on scattering theory and interactions. 
This work was supported by the Swiss NSF, MaNEP, the European Marie Curie MCRTN-CT-2003-504574 and the STREP project SUBTLE.


\section*{References}
\begin{thebibliography}{100}


\bibitem{Buettiker:93a} B\"uttiker M, Thomas H and  Pr\^etre A 1993
Mesoscopic capacitors {\it Phys. Lett.} A {\bf 180}, 364 

\bibitem{Gabelli} Gabelli J, Berroir J M, F\`eve G,
Pla{\c{c}}ais B, Jin Y, Etienne B and Glattli D C 2006
Violation of Kirchhoff's Laws for a Coherent RC Circuit
{\it Science} {\bf 313} 499

%J.~Gabelli, PhD dissertation, ENS Paris 6 (2006).

\bibitem{bil}
B\"{u}ttiker M, Imry Y and Landauer R 1983 Josephson Behavior in Small Normal One-Dimensional Rings {\it Phys. Lett.} A {\bf 96} 365


\bibitem{exp1}
Levy L P,  Dolan G, Dunsmuir J, and Bouchiat H, 1990 
Magnetization of mesoscopic copper rings: Evidence for persistent currents
{\it Phys. Rev. Lett.} {\bf 64} 2074

\bibitem{exp2}
Chandrasekhar V, Webb R A, Brady M J, Ketchen M B, Gallagher W J, and Kelinsasser A 1991
Magnetic response of a single, isolated gold loop
{\it Phys. Rev. Lett.} {\bf 67} 3578

\bibitem{exp3}
Mailly D, Chapelier C, and Benoit A 1993
Experimental observation of persistent currents in GaAs-AlGaAs single loop
{\it Phys. Rev. Lett.} {\bf 70}  2020



\bibitem{gia} Gefen Y, Imry Y, and Azbel M Ya 1984 Quantum oscillations and the Aharonov-Bohm effect for parallel resistors {\it Phys. Rev. Lett.} {\bf 52} 129 

\bibitem{bia} B\"uttiker M, Imry Y, and Azbel M Ya 1984 Quantum oscillations in one-dimensional normal-metal rings {\it Phys. Rev.} A {\bf 30}, 1982 

\bibitem{webb} Webb R A, Washburn S, Umbach C P, and Laibowitz R B 1985 Observation of h/e Aharonov-Bohm oscillations in normal metal rings  {\it Phys. Rev. Lett.} {\bf 54} 2696 

\bibitem{hansen} Hansen A E, Pedersen S, Kristensen A, Sorensen C B, Lindelof P E 2000
Investigation of the mesoscopic Aharonov-Bohm effect in low magnetic fields
{\it Physica} E {\bf 7} 776

\bibitem{qpc1}
van Wees B J van Houten H, Beenakker C W J, Williamson J G, Kouwenhoven L P, van der Marel D, and 
Foxon C T, 1988 Quantized conductance of point contacts in a two-dimensional electron gas {\it Phys. Rev. Lett.} {\bf 60}, 848 

\bibitem{qpc2} 
Wharam D A, Thornton T J, Newbury R, Pepper M, Ahmed H, Frost J E F, Hasko D G, Peacock D C, Ritchie D A and Jones G A 1988 One-dimensional transport and the quantization of the ballistic resistance {\it J. Phys.} C {\bf 21} L209 

\bibitem{butt88}
B\"uttiker M 1992 The Quantum Hall Effect in Open Conductors, in 
{\it Nanostructured Systems}, edited by Mark Reed,
Semiconductor and Semimetals,  Vol. 35, 191 (Academic Press, Boston). 

\bibitem{kouwen} 
Sohn L  L, Kouwenhoven L P and Schoen G, 1997
{\it Mesoscopic Electron Transport}, 
NATO Advanced Study Institute, Series E: Applied Science, Vol. 345. p. 259.
(Dordrecht: Kluwer Academic Publishers).

\bibitem{Buettiker:96a} Pr{\^e}tre A, Thomas H, and B{\"u}ttiker M 1996
Dynamic admittance of mesoscopic conductors: Discrete-potential model
{\it Phys. Rev.} B {\bf 54}, 8130 

\bibitem{pomorski} Pomorski P, Pastewka L, Roland C, Guo H, Wang J, 2004
Capacitance, induced charges, and bound states of biased carbon nanotube systems
{\it Phys. Rev.} B {\bf 69}, 115418  

\bibitem{Sillanpaa:05a} Sillanp\"a\"a M A, Lehtinen T, Paila A, Makhlin Yu, Roschier L and Hakonen P J 2005 
Direct Observation of Josephson Capacitance
{\it Phys. Rev. Lett.}
{\bf 95}, 206806 
  
\bibitem{Duty} Duty T, Johansson G, Bladh K, Gunnarsson D, Wilson C, and Delsing P 2005
Observation of Quantum Capacitance in the Cooper-Pair Transistor
{\it Phys. Rev. Lett.} {\bf 95}, 206807  

\bibitem{Flensberg:93a} Flensberg K 1993 
Capacitance and conductance of mesoscopic systems connected by quantum point contacts
{\it Phys. Rev.} B {\bf 48}, 11156
 
\bibitem{Matveev:95a} Matveev K A 1995
Coulomb blockade at almost perfect transmission 
{\it Phys. Rev.} B {\bf 51}, 1743

\bibitem{lehur1} Le Hur K 2004
Coulomb Blockade of a Noisy Metallic Box: A Realization of Bose-Fermi Kondo Models
{\it Phys. Rev. Lett.} {\bf 92} 196804

  
\bibitem{Buettiker:96b} 
B{\"u}ttiker M and Stafford C A 1996
Charge Transfer Induced Persistent Current and Capacitance Oscillations
{\it Phys. Rev. Lett.} {\bf 76}, 495 
         

\bibitem{gopar} Gopar V A, Mello P A, B\"uttiker M (1996) Mesoscopic Capacitors: A statistical analysis {\it Phys. Rev. Lett.} {\bf 77} 3005


\bibitem{brouwer1} Brouwer P and B\"uttiker M  1997
Charge-Relaxation and Dwell Time in the 
Fluctuating Admittance of a Chaotic Cavity 
{\it Europhys. Lett.} {\bf 37}, 441

\bibitem{brouwer2}
Brouwer P W, Frahm K M, and Beenakker C W J 1997
Quantum Mechanical Time-Delay Matrix in Chaotic Scattering
{\it Phys. Rev. Lett.} {\bf 78}, 4737 


\bibitem{nigg} Nigg S E, Lopez R and B\"uttiker M 2006 
               Mesoscopic charge relaxation
               {\it Phys. Rev. Lett.} {\bf 97}, 206804

\bibitem{wigner}
               Wigner E P 1952 On the connection between the distribution function of poles                      and residues for an R-function and its invariant derivative
               {\it Ann. Math} 55, 7 
           
\bibitem{Hackenbroich:01a} Hackenbroich G 2001 
{\it Phase coherent transmission through interacting mesoscopic systems} 
Phys. Rep. {\bf 343}, 463
 

\bibitem{rmt} Brouwer P W and Beenakker C W J 1997 
              Voltage probe and imaginary potential models for dephasing in a chaotic                           quantum dot
              {\it Phys. Rev.} B {\bf 55}  4695
              
              
\bibitem{Buttiker90} B\"uttiker M 1990 Quantized Transmission of a Saddle Point Constriction
              {\it Phys. Rev.} B {\bf 41} 7906         

\bibitem{mb87} B\"uttiker M 1987 Zero-current persistent potential drop across                                   small-capacitance Josephson junctions
               {\it Phys. Rev.} B 36, 3548

\bibitem{bouchiat} Bouchiat V, Vion D, Joyez P, Esteve D, Devoret M H 1998
                   Quantum Coherence with a Single Cooper Pair 
                   {\it Physica Scripta} {\bf T76}  165

\end{thebibliography}
\end{document}